\documentclass[]{aa}       
\usepackage{graphicx}
\usepackage{amssymb}
\usepackage{amsmath}
\usepackage{longtable}
\usepackage{supertabular}
\usepackage{natbib} 

\begin{document}

\title{The impact of red giant/AGB winds on AGN jet propagation}

\author{M. Perucho\inst{1,2} \and
        V. Bosch-Ramon\inst{3} \and
        M. V. Barkov \inst{4,5,6}
    }

\authorrunning{Perucho, Bosch-Ramon, Barkov}

\titlerunning{Red giant/AGB winds and AGN jets.}

\institute{Dept. d'Astronomia i Astrof\'{\i}sica, Universitat de Val\`encia, C/ Dr. Moliner 50, 46100, Burjassot (Val\`encia), Spain; 
Manel.Perucho@uv.es 
\and
Observatori Astron\`omic, Universitat de Val\`encia, C/ Catedr\`atic Jos\'e Beltr\'an 2, 46980, Paterna (Val\`encia), Spain
\and
Departament de F\'{i}sica Qu\`antica i Astrof\'{i}sica, Institut de Ci\`encies del Cosmos (ICCUB), Universitat de Barcelona, IEEC-UB, Mart\'{i} i Franqu\`es 1, E08028 Barcelona, Spain
\and
Astrophysical Big Bang Laboratory, RIKEN, Saitama 351-0198, Japan
\and
DESY, Platanenallee 6, 15738 Zeuthen, Germany
\and
Department of Physics and Astronomy, Purdue University, 525 Northwestern Avenue, West Lafayette, IN 47907-2036, USA
}

\offprints{M. Perucho, \email{manel.perucho@uv.es}}

\date{Received <date> / Accepted <date>}

\abstract
{Dense stellar winds may mass-load the jets of active galactic nuclei, although it is unclear what are the time and spatial scales in which the mixing takes place.}  
{Our aim is to study the first steps of the interaction between jets and stellar winds, and also the scales at which the stellar wind may mix with the jet and mass-load it.}
{We present a detailed two-dimensional simulation, including thermal cooling, of a bubble formed by the wind of a star, designed to study the initial stages of jet-star interaction. We also study the first interaction of the wind bubble with the jet using a three-dimensional simulation in which the star enters the jet. Stability analysis is carried out for the shocked wind structure, to evaluate the distances over which the jet-dragged wind, which forms a tail, can propagate without mixing with the jet flow.}
{{ The two-dimensional simulations point at quick wind bubble expansion and fragmentation after about one bubble shock crossing time. Three-dimensional simulations and stability analysis point at local mixing in the case of strong perturbations and relatively small density ratios between the jet and the jet dragged-wind, and to a possibly more stable shocked wind structure at the phase of maximum tail mass flux. Analytical estimates also indicate that very early stages of the star jet-penetration time may be also relevant for mass loading.} The combination of these and previous results from the literature suggests highly unstable interaction structures and efficient wind-jet flow mixing on the scale of the jet interaction height.}
{The winds of stars with strong mass-loss can efficiently mix with jets from active galactic nuclei. In addition, the initial wind bubble shocked by the jet leads to a transient, large interaction surface. The interaction between jets and stars can produce strong inhomogeneities within the jet. {As mixing is expected to be effective on large scales, even individual AGB stars can significantly contribute to the mass-load of the jet and thus affect its dynamics. Shear layer mass-entrainment could be important.} The interaction structure can be a source of significant non-thermal emission.} 
\keywords{Galaxies: active -- Galaxies: jets -- relativistic processes -- shock waves}

\maketitle

\section{Introduction} \label{intro}

Jets of active galactic nuclei (AGN) are collimated outflows originated at the core of galaxies, very likely in the vicinity of a  supermassive black hole (SMBH) that accretes matter from its environment \citep[e.g.][]{beg84}. Among the AGN population, the well-known morphological dichotomy between FRI and FRII radiogalaxies \citep{fr74} has been claimed to be related to jet power \citep[e.g.,][]{rs91}. On the one hand, the extragalactic FRII jets, with powers typically over $10^{44}$~erg~s$^{-1}$, show an edge-brightened structure in radio. In this case, the jets show a remarkable stability and reach the interaction site with the ambient still collimated, generating a strong forward bow-shock, and a reverse shock, known as hot-spot \citep[e.g., Cyg~A,][]{cb96,mk16}. On the other hand, the edge-darkened jets in radio are related to lower powers and, although they may show a collimated morphology on parsec scales, there is a transition to large opening angles or decollimation (the flaring zone) along the first kiloparsecs that leads to the absence of a hot-spot at the ambient interaction regions \citep[e.g., 3C~31,][]{lb02a}. The suggested reason for the observed jet deceleration has been typically claimed to be the entrainment process of cold, slow gas that is incorporated and mixed with the jet flow \citep{bi84,la96}.

Mass entrainment seems a natural outcome of jet propagation. Within their first kiloparsecs, jets evolve inside their host galaxies, which contain large amounts of gas, dust, and stars, mainly close to the nucleus \citep[e.g.,][]{bur70}, so there is plenty of material the jet could interact with. It seems therefore unavoidable that stars and medium inhomogeneities will frequently interact with, and in some cases penetrate into, the AGN innermost jet regions. In the case of FRI jets, two main processes have been invoked to explain entrainment: (i) mixing in a turbulent shear layer between the jet and the ambient \citep[e.g.][]{dy86,dy93,bi94,wa09}, and (ii) injection from stellar mass loss \citep[e.g.][]{ph83,ko94,blk96,hub06}.

\citet{lb02b} constructed a model of the jet in 3C~31 using the basic conservation laws and the velocity field inferred by \citet{lb02a}, and concluded that the continuous deceleration in the jet requires a monotonic increase of the entrainment rate at large distances, which is incompatible with mass-load from stars, whose density falls rapidly with distance, so entrainment from the galactic atmosphere across the boundary layer of the jet was favored far from the nucleus. Closer to the active nucleus, mass-load by stars could still be relevant. Concerning the process of entrainment through a turbulent shear layer, \citet{PM07} showed, via a two-dimensional (2D) axisymmetric simulation, that a recollimation shock in a light jet formed in reaction to steep interstellar density and pressure gradients may trigger large-scale nonlinear perturbations that lead to jet disruption and mixing with the external medium.

Recent work by \citet{lb14} has shown that continuous deceleration of FRI jets by small-scale instabilities at the shear layer could better explain the jet properties inferred from observations and modeling downstream of the flaring region, where these jets increase their brightness and also start to decollimate. This is in agreement with \citet{bi84} and \citet{wa09}, who developed models based on this idea. Numerical simulations of the development of Kelvin-Helmholtz instabilities in 2D and 3D also support the slow and continuous entrainment and deceleration due to the growth of short-wavelength Kelvin-Helmholtz instability modes \citep{pe05,pe10}. The relative irrelevance of mass-load by stellar winds from an old population of stars (as expected in giant elliptical galaxies typically hosting jets) on large scales has been confirmed by numerical simulations \citep{pmlh14}.

On the one hand, mass-load could still represent an efficient deceleration mechanism in the case of low-power jets ($L_j\leq 10^{42}$~erg~s$^{-1}$) \citep{ko94,pmlh14}. Following the conclusions from \citet{lb14}, it remains possible that the weaker sources of the sample of ten FRI~jets in \cite{lb14}, namely M84 and NGC193, are decelerated primarily by stellar mass-loading. On the other hand, even a single star with large mass-loss rates may affect significantly the dynamics of relatively weak jets. \citet{hub06} concluded that a Wolf-Rayet star can temporally quench the whole jet flow for jet powers $L_j\leq 10^{42}$~erg~s$^{-1}$. \citet{hu13} have shown that a star with a powerful wind ($\dot{M} = 10^{-4}$~M$_\odot$~yr$^{-1}$) can produce observable structures in jets with relatively low powers ($L_j=2.4\times10^{43}$~erg~s$^{-1}$).

Regarding the powerful jets of FRII AGN, observations reveal a decrease in the jet flow velocity from parsec, where Lorentz factors are of the order of tens \citep[see, e.g.,][]{lis13}, to kiloparsec scales, where Lorentz factors are closer to 1 \citep{mh09}. This deceleration could also be partly caused by stellar wind mass-load, although it is more probable that it is due to mass-load at the shear-layer and to the conversion of kinetic energy into internal energy at conical shocks along the jet.

As noted above, however, jet interaction with clouds and stars is still expected given the richness of the AGN environment. Even if these objects play a minor role in the jet evolution on kpc scales, this does not prevent a significant dynamical and radiative impact of their interactions with the jet in other contexts. For instance, stellar wind and cloud mass-load in the first parsec may explain the transition from pair- to proton-dominated jets \cite[see][and references therein]{brpb12,kh13}. In addition, interaction with stars and clouds has been claimed to explain the presence of knots in M87 \citep{bk79,cb85}, and in the jet of Centaurus~A; knots detected at tens to hundreds of parsecs away from the source could correspond to this kind of interactions \citep{wo08,go10}. More recently, VLBI observations of the jet in Centaurus~A have also revealed jet structure interpreted as possible jet-star interaction at sub-parsec scales \citep{mu14}. Also, mass-loading by stellar winds has been recently claimed to explain the pressure imbalance between the lobes and the ambient medium in Centaurus~A (\citealt{wy13}; see also \citealt{wy15} and references therein). Finally, the radiative counterpart of the interaction of relativistic flows with stellar winds or clouds may be important as well, due to the strong shock produced and the subsequent particle acceleration, which could be, at least in some cases, responsible for the production of steady and variable gamma-ray emission in AGN jets \citep{bp97,bar10,ba12b,brpb12,kh13,ara13,bos15,bb15,del16,bbs16,abk17}. Therefore, in short, entrainment by stellar winds cannot be neglected for the study of jet content, long-term jet and lobe evolution, and even to explain emission features at high energies.

The global impact of stars or clouds on the jet propagation and content has been studied analytically or numerically in 
the past by different authors \citep[e.g.][]{ko94,blk96,ste97,cho05,hub06,sut07,jey09}. More specific analytical 
calculations of the dynamical interaction between one cloud or red giant (RG) with an AGN jet have been carried out in 
the context of radiation studies \citep[e.g.][]{ara10,bar10,ba12a,kh13}, whereas \cite{brpb12} performed relativistic 
hydrodynamic simulations of the interaction of a RG in the inner-most region of the jet. In the latter, the star was simulated 
as a high-density core, surrounded by an atmosphere with a radial profile in density, and in pressure equilibrium. 
The main results were the formation of a powerful shock, potential site of non-thermal processes, and the expansion, 
disruption and advection of the material ablated by the jet. The density profile of the wind showed to play an important 
role: whereas homogeneous clouds impacted by the jet quickly expand, a $r^{-2}$-density profile leads to a smooth 
atmosphere ablation \citep[with some dependence of the mass extraction rate on the numerical resolution,][]{brpb12}. 
In jet regions located farther downstream, it is the wind of the star and not its atmosphere what interacts with 
the jet, and the shocked wind density profile will determine the properties of the jet-wind interaction at different stages of the process: (i) At first, the star carries a wind bubble in (ram) pressure equilibrium with its environment; simulated here in 3D and treated analytically in the discussion. (ii) Then, right after the star has entered the jet, there is a transitory stage corresponding to the interaction between the jet and the stellar wind bubble that has survived jet penetration (this paper concentrates on phases one and two). Finally, (iii) there is a third stage, the steady wind-jet interaction; phase three, not studied here (see \citealt{del16}). Although they may not be the most relevant regarding mass-load, the first two stages imply a sudden quick release of mass into the jet, and a potential large target for jet interaction, energy dissipation and non-thermal emission, hence the importance of its characterization.

In this paper, we focus on the first stage of the jet-wind interaction, and present relativistic hydrodynamical simulations of a RG/AGB (asymptotic giant branch) star surrounded by its wind interacting with a relativistic jet with moderate power, at 100~pc from the galactic nucleus. We assume that the RG/AGB star has just entered into the jet, and is releasing the wind bubble formed prior to the entrance in the jet. The initial stages of the process involve the interaction of the jet with the wind bubble. The high density and compression of the gas in the bubble anticipate that cooling must be included, and it represents an important point of this work. Taking into account that the formation of a cometary tail is a natural outcome of the interaction in parallel with the dispersion of the wind bubble, we also study the typical disruption scales of the tail using linear Kelvin-Helmholtz instability analysis. In other words, we study the wind bubble evolution inside the jet, and the formation and properties of the cometary-like tail, which is also present during the interaction. This tail is eventually incorporated to the jet flow via the growth of instabilities, giving an approximate distance at which this gas could be completely mixed with the jet flow for the case of the formation of a stable channel of shocked stellar wind. The aim is to estimate the scales at which all these processes occur, extending and complementing previous analytic \citep[e.g.][]{ara10,bar10,ba12a,kh13} and numerical \citep{brpb12,pmlh14,bos15,del16} studies of this scenario performed by the authors \citep[see also, e.g.,][]{hub06}.

The points described in the previous paragraph have been investigated using numerical simulations. The high resolution and simulation time required to follow the evolution of this thin layer of dense, cold gas makes to run such a simulation in 3D unfeasible. Therefore, we have run a 2D axisymmetric simulation to study the physical scenario in as much detail as possible. This decision has a clear negative effect in terms of the evolution of the cometary tail and mixing, which is limited in the 2D case by the lack of development of helical instability modes and turbulence. Nevertheless, this caveat does not affect (i) the amount of gas that will eventually be mass loaded, as the latter is the amount of gas in the stellar wind envelope, nor (ii) the basic dynamics of the bubble. In addition, we have performed a short 3D simulation for the initial stages of the interaction of a star with the jet, during its first interaction with the jet boundary. At this stage, we neglect the role of the magnetic field in the jet, accounting only for its ram pressure. Relativistic magnetohydrodynamic simulations of this scenario will be performed in our future work. We emphasize that we do not investigate here the absolute jet mass-load by stellar winds, but rather the evolution of the shocked wind in a single jet-star interaction, and the possible distance scales of tail-jet mixing.

The paper is structured as follows: in Section~\ref{sim} we introduce the physical and the numerical setup of the simulations; in Sect.~\ref{res} we present the results of our simulations; Sect.~\ref{disc} is devoted to the discussion of our results, and in Sect.~\ref{sum} we summarize our work.

\section{Simulations}\label{sim}

   For the simulations, we used the finite-volume code {\it Ratpenat}, which solves the equations of relativistic hydrodynamics in conservation form using high-resolution-shock-capturing methods. {\it Ratpenat} is a hybrid parallel code  -- MPI + OpenMP -- \citep[e.g.,][]{pe10}. For visualization we used IDL software and LLNL VisIt \citep{visit}. 
   
The conservation equations for a relativistic flow in two-dimensional cylindrical coordinates ($R,\, z$), assuming
axisymmetry and using units in which $c=1$, are:
\begin{equation}
  \frac{\partial \mathbf{U}}{\partial t} + \frac{1}{R}\frac{\partial R
\mathbf{F}^R}{\partial R} + \frac{\partial \mathbf{F}^z}{\partial z} =
\mathbf{\Sigma} ,
\end{equation}
with the vector of variables 
\begin{equation}
  \mathbf{U}=(D,D_{\rm l},S^R,S^z,\tau)^T ,
\end{equation}
the vector of fluxes
\begin{equation}
  \mathbf{F}^R=(D v^R , D_{\rm l} v^R , S^R v^R + p , S^z v^R , S^R - D v^R)^T ,
\end{equation}
\begin{equation}
  \mathbf{F}^z=(D v^z , D_{\rm l} v^z , S^R v^z , S^z v^z + p, S^z -D v^z)^T ,
\end{equation}
{\bf and the source terms
\begin{equation} \label{eq:source}
  \mathbf{\Sigma} =  (0, -\Lambda u^1, -\Lambda u^2, -\Lambda u^3, -\Lambda u^0)^T ,
\end{equation}
 with $u^\mu$ ($\mu=0,1,2,3$) being the natural velocities.}
 The five unknowns $D,D_{\rm l},S^R,S^z$ and $\tau$, stand for the densities of
the total and leptonic rest masses, the radial and axial components of the momentum, and the energy
(excluding the rest mass energy), respectively. All five variables are defined in the laboratory
frame, and are related to the quantities in the local rest frame of the
fluid (primitive variables) according to:
\begin{equation}
  D = \rho W,
\end{equation}
\begin{equation}
  D_{\rm l} = \rho_{\rm l} W,
\end{equation}
\begin{equation}
  S^{R,z} = \rho h W^2 v^{R,z},
\end{equation}
\begin{equation}
  \tau=\rho h W^2\,-\,p\,-\,D,
\end{equation}
where $\rho$ and $\rho_{\rm l}$ are the total and the leptonic rest-mass
densities, respectively, $v^{R, z}$ are the components of the velocity
of the fluid. $W$ is the Lorentz factor ($W = (1-v^i v_i)^{-1/2}$, with summation over repeated indices), and $h$ is the specific
enthalpy defined as
\begin{equation}
  h = 1 + \varepsilon + p/\rho,
\end{equation}
where $\varepsilon$ is the specific internal energy and $p$ is the
pressure. The system is closed by the Synge equation of state
\citep{sy56}, as described in Appendix~A of \citet{PM07}.  This equation of state accounts for a mixture
of relativistic Boltzmann gases (in our case, electrons, positrons and protons). The code also integrates an equation for the jet mass fraction, $f$. This quantity, set to 1 for the injected jet material and 0 otherwise, is used as a tracer of the jet material through the grid. Cooling has been introduced as a source term, $\Lambda$, in the energy equation. The term is defined (in cgs) units as in the approximation given by \cite{mya98}:

\begin{equation}
 \Lambda\,=\, n_e\,n_Z \times
 \left\{ 
 \begin{array}{lr}
 7\times10^{-27}T,& 10^4\leq T \leq 10^5  \\  \nonumber
 7\times10^{-19}T^{-0.6},& 10^5\leq T \leq 4\times10^7 \\
 3\times10^{-27}T^{0.5},& T \geq 4\times10^7    \nonumber
 \end{array} \right .
\end{equation}

Here, we have taken $n_e = \rho_e/m_e$ and $n_Z=n_p =\rho_p/m_p$ ($Z=1$, we assume hydrogen for simplicity). The equation of state that we use allows us to account for the number of electrons or pairs in a given cell, with $\rho_e$ and $\rho_p$ the electron and proton densities, respectively. The cooling fraction is thus computed for each cell using the described procedure. We note in this respect, that the particles in each cell are assumed to be in thermodynamic equilibrium and contribute to the local temperature as described in \citep{sy56}. We consider that the thermal radiative losses are only relevant for the denser, non-relativistic wind in our simulation, where $v\ll c$ (the shocked wind velocity is $v_w=10^{-4}\,c$, and the shocked wind velocity at the dense regions is even smaller, see below). Therefore, we consider that these losses are isotropically emitted and that $u^0=1$, $u^i = 0$ in Eq.~\ref{eq:source}.

 The 3D simulation was performed solving the RHD equations in cartesian coordinates. In this case, no cooling source terms were included, and an ideal gas equation of state was used, i.e., a single population of particles was considered.

\subsection{Two-dimensional, radiative, axisymmetric simulation (SW2c)}
In our 2D simulation, we focus on the initial, transitory phase of interaction between a stellar wind envelope (bubble) and a jet at $\sim$100~pc from the central black hole. In particular, we study the effects of cooling and the presence of a dense shell of shocked wind gas surrounding the stellar wind bubble. The unit distance of the simulation is the size of the injector $R_0 = 1.1 \times 10^{16}~{\rm cm}$, and the unit density is the density given to the wind at the injection region, $\rho_{w,0}=1.5 \times 10^{-19}~{\rm g~cm^{-3}}$. The grid is formed by a homogeneous resolution region involving a domain in cylindrical coordinates $(r,\,z)$ of $[0,\,100]\,R_0\, \times [0,\,200]\, R_0$ ($1.1\times 10^{18}\,{\rm cm}$ $\times$ $2.2\times 10^{18}\,{\rm cm}$), with a resolution of 6.4 cells per $R_0$, adding up a total of $640 \times 1280$ cells.

  \begin{figure}[!t]
  \includegraphics[clip,angle=0,width=\columnwidth]{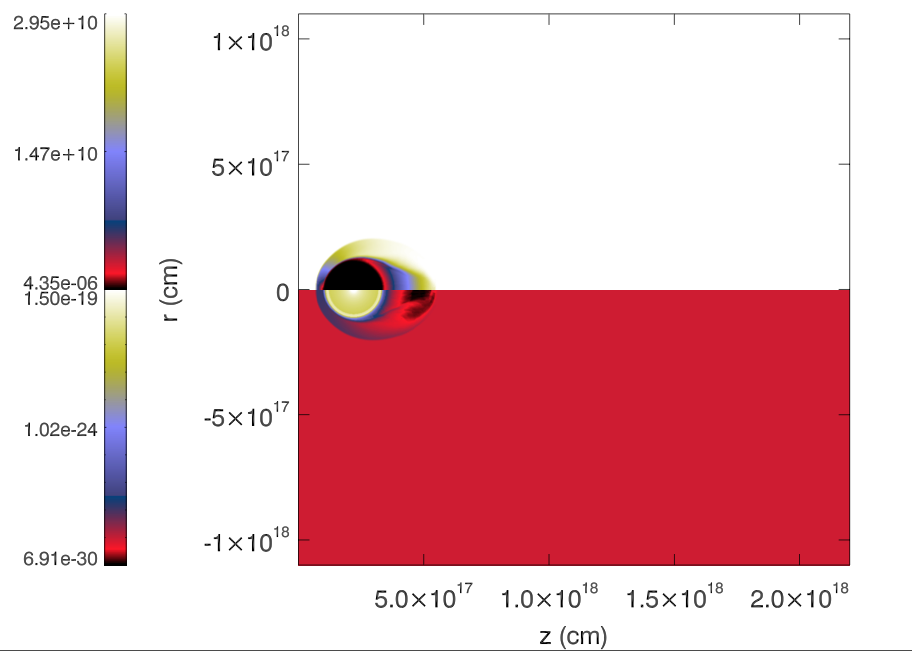}\\
  \caption{Snapshot of rest-mass density (bottom half) and modulus of velocity (top half) of simulation SW2c at $t=7.4\times10^6$~s, in which we make profit of the axisymmetric nature of the simulation. The plot only includes the central, homogeneous grid to show the setup of the simulation.}
  \label{fig:SW2a}
  \end{figure} 

Figure~\ref{fig:SW2a} shows a snapshot of the rest-mass density during the initial stages of simulation SW2c to illustrate the setup of the simulation. The numerical grid is filled by the jet flow, but for a circular region occupied by the wind injector, the bubble, and a dense shell of shocked wind gas expected for a supersonic wind. The injector, at the star location, is simulated as the cells within 1~$R_0$, where the properties of the flow are fixed as a boundary condition. It is located at position $(0,20~R_0)$, i.e., on axis, at $20\,R_0$ ($z=2.2\times 10^{17}$~cm) from the simulation boundary in which the jet flow is injected. The wind is injected with radial velocity $v_w=10^{-4}\,c$. The stellar wind parameters result in a mass-loss rate $\dot{M}_w=10^{-5}$~M$_\odot$~yr$^{-1}$. The boundary conditions are reflection on the jet axis ($r=0\,R_0$), (jet) injection at $z=0\,R_0$  (left boundary as seen in Fig.~\ref{fig:SW2a}) and outflow at $r=100\,R_0=1.1\times 10 ^{18}\,{\rm cm}$ and $z=200\,R_0=2.2\times 10 ^{18}\,{\rm cm}$ (top and right in the figure, respectively).

 The setup of the 2D axisymmetric simulation requires that the wind injector is located at the symmetry axis. This is a simplification of the scenario, because we locate the wind bubble inside the jet from the beginning of the simulation, i.e., we do not follow the process of entrance of the star into the jet. This also implies that the dense shell of shocked wind gas that surrounds the jet remains basically untouched until the star is fully inside the jet (this is relaxed in our 3D simulation (see next section), which was designed to study the interaction of a large bubble of stellar wind formed outside the jet, with the jet boundary).

An initial region with size 10~$R_0$, the bubble, is defined around the injector. Such an initial bubble size implies a penetration velocity 10 times higher than $v_w$. The bubble region is filled with a wind with decreasing density $\rho_w(r) = \rho_{w,0}(R_0/r)^2$ and an initial (constant) temperature $T_w = 10^3~{\rm K}$, which is dynamically irrelevant. A dense shell of shocked wind material is set surrounding the wind bubble: the density of the bubble at its outer-most 10\% radial region is taken 10 times higher than that obtained from a pure $r^{-2}$-density profile, to mimic the presence of cooled shocked wind at its termination from the beginning of the simulation. 

The presence of such a shell, formed by shocked wind gas that surrounds the wind bubble, is probably an important ingredient for the mass-load of the jet during the first interaction phase. This shell can be dense enough to change the mass-load pattern in the long term along this phase. Furthermore, thermal cooling of the wind should be considered in this case because it can be relevant for the evolution of the shocked shell. 

We used a version of the code that includes the Synge equation of state \citep{sy56} with two populations of particles, namely, leptons (electrons and positrons) and baryons (protons). This allowed us to set up the jet as composed of pairs, i.e., thermally hot, and the simulated stellar wind as a proton-electron flow. Furthermore, we used thermal cooling terms, following the approximation used in \cite{mya98} to account for the cooling of the dense, lower-temperature wind and shocked-wind gas. 
The jet conditions have been chosen to obtain a jet power $L_j\sim 10^{44}$~erg~s$^{-1}$ choosing the jet radius to be $R_j=10\,{\rm pc}$, i.e., we consider the interaction to take place at $\sim$100~pc from the central black hole for a jet with opening angle of $\xi=0.1$~radians. The rest-mass density of the pairs that form the jet is $\rho_j = 6.3 \times 10^{-29}$~g~cm$^{-3}$, the jet velocity $v_j=0.9798\,c$ (corresponding to a Lorentz factor $\gamma_j = 5$), the temperature $T_j=10^9~{\rm K}$, and the pressure $P_j=10^{-8}$~dyn~cm$^{-2}$. 

The simulation was run in Tirant, at the University of Val\`encia, in 128 cores, taking approximately $8\times 10^5$~computing hours, mainly owing to the necessary cooling-time check at the time-step calculation routine. The dynamical time-scales of the whole jet-bubble interaction process simulated is given by the scale of the bubble divided by the wind velocity, i.e., $\sim 10^{10}\,{\rm s}$. With the resolution used, each simulation time-step represents $\simeq 0.045\,R_0/c \times 3.67\times10^5 \,s /(R_0/c) ~ 1.65\times 10^4$~seconds. Therefore, the simulation requires $\simeq 10^6$ code steps to be completed, which is very demanding in terms of computing time, taking into account that each time-step took $\approx 25\,s$ using 128 cores (this means that it would take around 1 hour for a single core to make one step). Altogether, the complex thermodynamics used and the time-scale of the evolution of the system convert this simulation into a challenging approach to the scenario, which is currently impossible to reproduce in 3D with the same resolution and similar simulation runtimes. This would multiply the time needed to run the simulation by a factor equal to the number of cells used in the third dimension, i.e., a factor larger than one hundred.
In summary, this simulation offers a good approach to the physics of the system, because, even if the setup and geometry miss 3D effects, the computing effort and its emphasis are focused on the (thermo-)dynamics of the interaction.

It is relevant to note that after the first stage is passed, in which this bubble is shocked and dragged away by the jet, only a central region of the injector remains. In a real scenario, the size of this region is determined by jet-wind ram pressure equilibrium, as long as the star is still within the jet cross-section. This means that, once the wind bubble is gone, our simulation does not properly resolve the shocked jet-wind structure in the second, steady, stage of the interaction, because this stage occurs within our injection boundary condition, and it is therefore not considered here.

\subsection{Three-dimensional simulation (SW3)}

As a complement for our 2D simulation, we run a short, 3D simulation of a RG/AGB star wind bubble penetrating a relativistic jet (simulation SW3), focusing on the entrance of the star into the jet. This simulation was performed at Mare Nostrum, at the Barcelona Supercomputing Centre, using 256 cores during $5\times10^5$~computing hours.

In 3D, the injector is simulated as a group of cells within 1~$R_0$, in which the properties of the flow are fixed as a boundary condition at a given instant. In contrast with the 2D simulation, the injector propagates across the grid, entering the jet flow from its surrounding medium.

The grid is formed by the domain, in the $x,\,y,\,z$ cartesian coordinates, $[-10,\,10]\,R_0\, \times [0,\,240] \,R_0\times [0,\,160]\,R_0$, involving a total of $64 \times 768 \times 512$ cells. The grid is divided in two regions: In the region $y=[0,\, 2.7\times 10^{17}]\,{\rm cm}$ (i.e., $[0,\,25] \,R_0$), we set up the ambient medium (see, e.g., Figs.~\ref{fig:SW3a} and \ref{fig:SW3c}), in which the initial bubble and injector are embedded. The jet occupies the region $y> 2.7\times 10^{17}\,{\rm cm}$, and is set to flow in the positive $z$ direction (upwards in Figs.~\ref{fig:SW3a} and \ref{fig:SW3c}). The separation between the jet and the ambient medium is thus located at $y=25\,R_0$, along the $x$ and $z$ axis. In order to avoid numerical noise, a shear layer has been introduced between the ambient and the jet \citep[see, e.g.][]{pe04}. The shear layer has a width $\simeq 12\,R_0$ (extending $6\,R_0$ around $y=25\,R_0$). The wind injector is placed at $(0,\,13,\,20)$ in $R_0$ units, at $t=0$, i.e., at a distance $\simeq 12\,R_0$ from the shear-layer. The wind region occupies a spherical region in the grid with 64 cells of radius within this region. 
    
The sizes of the bubble and the whole simulated region are small enough with respect to the size of the jet ($2\times10^{17}\,{\rm cm}\,-\, 1.5\times10^{18}\,{\rm cm}$ vs $6\times 10^{19}\,{\rm cm}$) that we can approximate the simulated jet region to a slab, thus omitting the curvature of the jet surface. The boundary conditions are (jet) injection at $z\,=\,0\,R_0$ and outflow at the other five boundaries of the simulation box. The time-step of the simulation is $\simeq 0.075\,R_0/c = 2.75\times10^5\,s$. 


The setup of this simulation is simplified with respect to the simulation SW2c. We use a single ideal gas with adiabatic exponent of $5/3$ for the whole grid. This implicitly imposes a jet dominated by a non-relativistic proton gas, but this does not significantly affect the results as we focus on the non-relativistic bubble, and largely simplifies and speeds-up our calculations. In addition, we note that, as the jet is strongly supersonic, the nature of its content does not significantly affect the results obtained from this setup. The jet conditions have been also chosen to obtain a jet power $L_j\sim 10^{44}$~erg~s$^{-1}$. with a jet radius of $R_j=10\,{\rm pc}$, as for simulation SW2c. Therefore, the rest-mass density of the jet gas is $\rho_j = 6.3 \times 10^{-29}$~g~cm$^{-3}$, the jet velocity $v_j=0.9798\,c$ (corresponding to a Lorentz factor $\gamma_j = 5$), the temperature $T_j=10^9~{\rm K}$, and the pressure $P_j=10^{-8}$~dyn~cm$^{-2}$.

The medium pressure is set in equilibrium with the jet thermal pressure in order to avoid transversal motions, resulting in a temperature $T\simeq 10^3\,{\rm K}$ and a density $\rho_a=4.53\times10^{-23}\,{\rm g~cm^{-3}}$. An initial region with size 10~$R_0$, the bubble, is defined around the injector, with a wind with decreasing density $\rho_w(r) = \rho_{w,0}(R_0/r)^2$ and an initial (constant) temperature $T_w = 10^3~{\rm K}$, which is dynamically irrelevant. The initial size of the bubble is chosen to be in equilibrium with the thermal pressure of the ambient medium, which is one hundredth of the jet ram pressure. Such parameters have been chosen to emphasize the bubble being shocked and disrupted by the jet ram pressure. In this case, the injector approaches the jet with a velocity $v_k=1.1\times 10^7$~cm~s$^{-1}$, which would be the Keplerian velocity for a black hole mass of $M_{BH}\simeq 3\times 10^8$~M$_\odot$ at a distance of 100~pc. The initial distance between the injector and the shear layer and this velocity give a time $\simeq 10^{10}\,s$ for the beginning of the interaction.     

\section{Results}\label{res}

\subsection{SW2c}\label{resc}

Figure~\ref{fig:SW2a} shows that the jet-wind interaction evolves as expected during the initial stages of simulation, with the formation of a bow-shaped shock in the jet flow within a short time. Ideally, the jet and wind properties give a stagnation point at the equilibrium position given by:
\begin{equation}
R_s=\sqrt{\frac{\dot{M}_w\,v_w}{4 \pi\,\rho_j\,\gamma^2\,v_j^2 }}\simeq 1.1 \times 10^{16}\,{\rm cm}
\end{equation}  
from the star (i.e. $R_s\simeq R_0$). In reality, the shock propagates through the shell and the initial wind bubble\footnote{See \cite{kmc94} and \cite{pfb02} for a detailed discussion on the hydrodynamics of the interaction between shocks and clouds.} and stops not far from the injector after a time $t\simeq 1.5\times 10^{10}\,{\rm s}$.

   \begin{figure}[h]
   \includegraphics[clip,angle=0,width=\columnwidth]{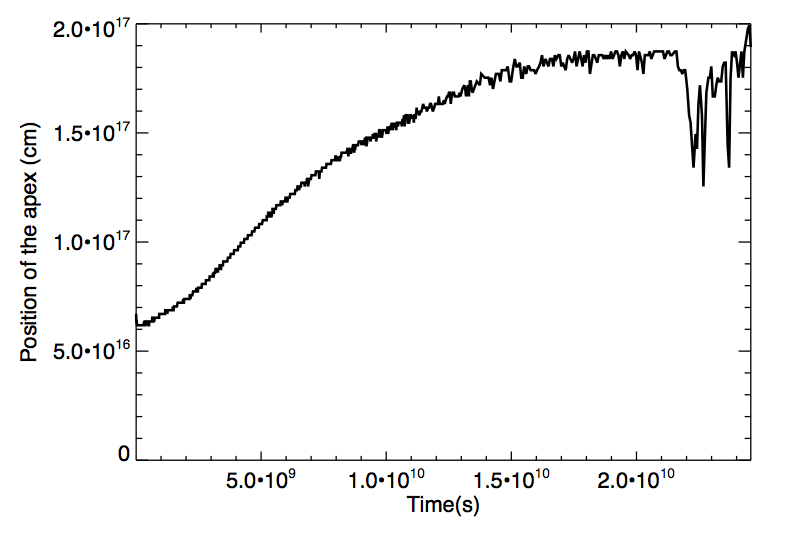}
     \caption{Position of the head of the shock in the jet flow along the axis in simulation SW2c, with the origin set at the left corner of the grid. The centre of the injector is located at $z=2.2\times10^{17}~{\rm cm}$.}
   \label{fig:apex3}
   \end{figure}

Fig.~\ref{fig:apex3} shows the axial position of the shock produced in the jet with time. It stabilizes at $z=1.8-1.9\times10^{17}~{\rm cm}$; i.e. $r=3-4\times10^{16}~{\rm cm}$ from the star. This is about three times $R_s$, significantly farther than it should be. The reason for this is partially that the steady state jet-wind interaction region is not resolved, and the propagation of the stagnation point towards the star is stopped when it approaches the injector boundary condition due to limited resolution: at $r=3-4\times10^{16}~{\rm cm}$ from the star, the number of cells between this position and the injector is only $\simeq 20$. We note that in a jet-wind interaction, $R_s$ is approximately the distance from the star to the wind termination shock, and not to the shock in the jet, but this could hardly explain a jet shock position at $\sim 3\,R_s$. Therefore, the simulation fails to properly describe de jet-wind interaction close to the star for $t \gtrsim 1.5 \times 10^{10}$~s. Nevertheless, the evolution of the shocked initial wind bubble is still properly described up to later times far from the star. 

\begin{figure*}[!t]
  \includegraphics[clip,angle=0,width=0.96\columnwidth]{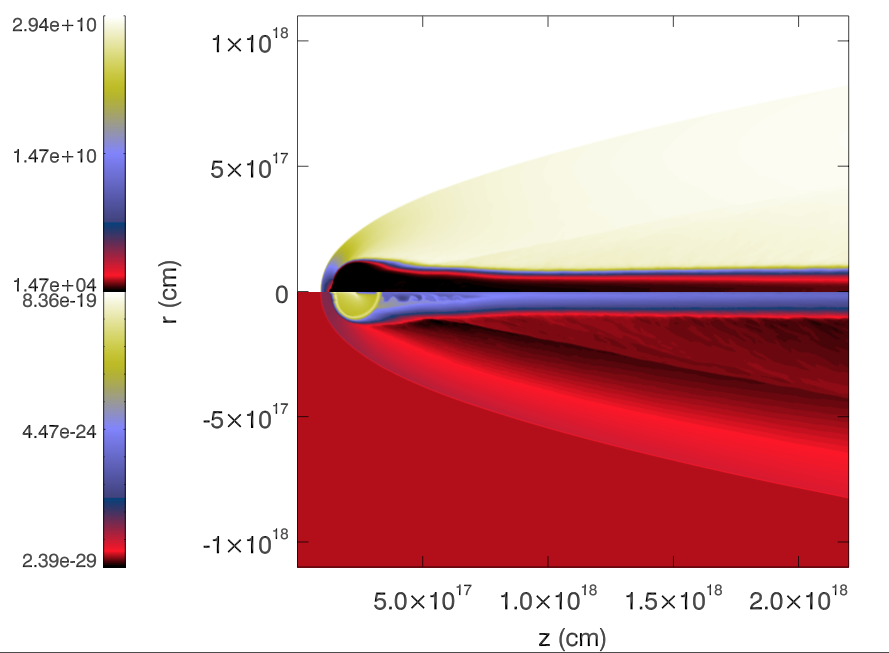}\,  
  \includegraphics[clip,angle=0,width=0.96\columnwidth]{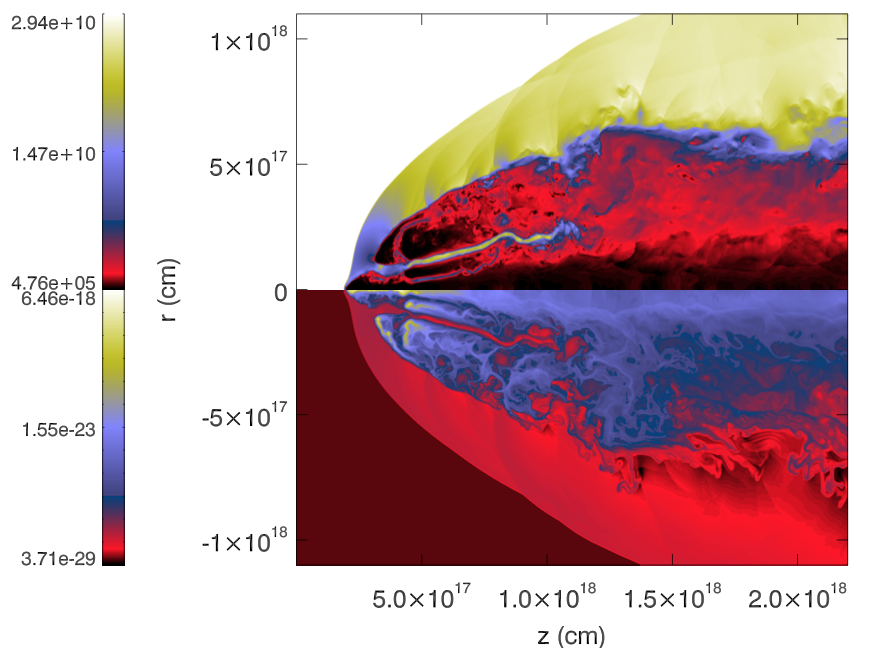}\\
   \includegraphics[clip,angle=0,width=0.96\columnwidth]{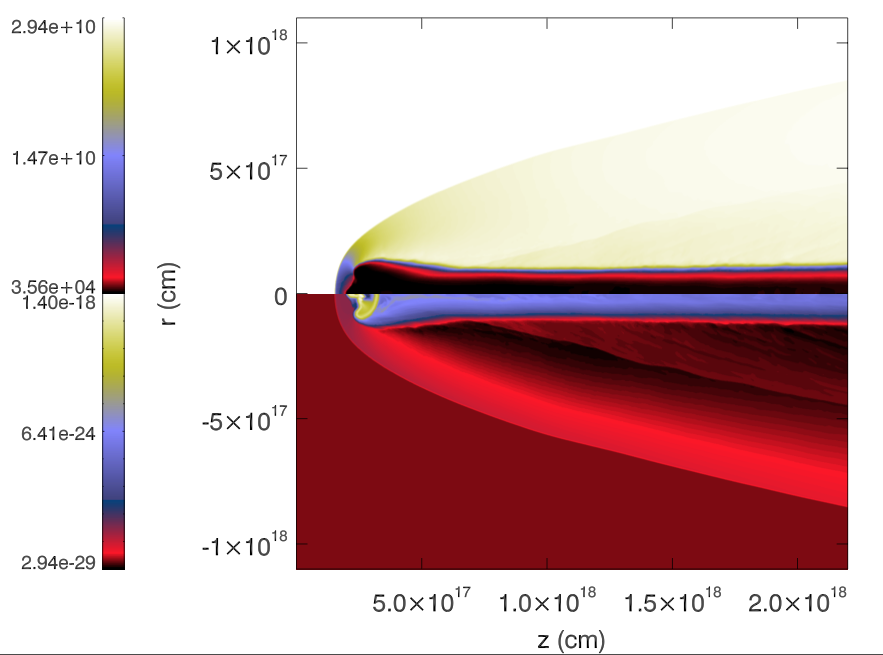}\,
   \includegraphics[clip,angle=0,width=0.96\columnwidth]{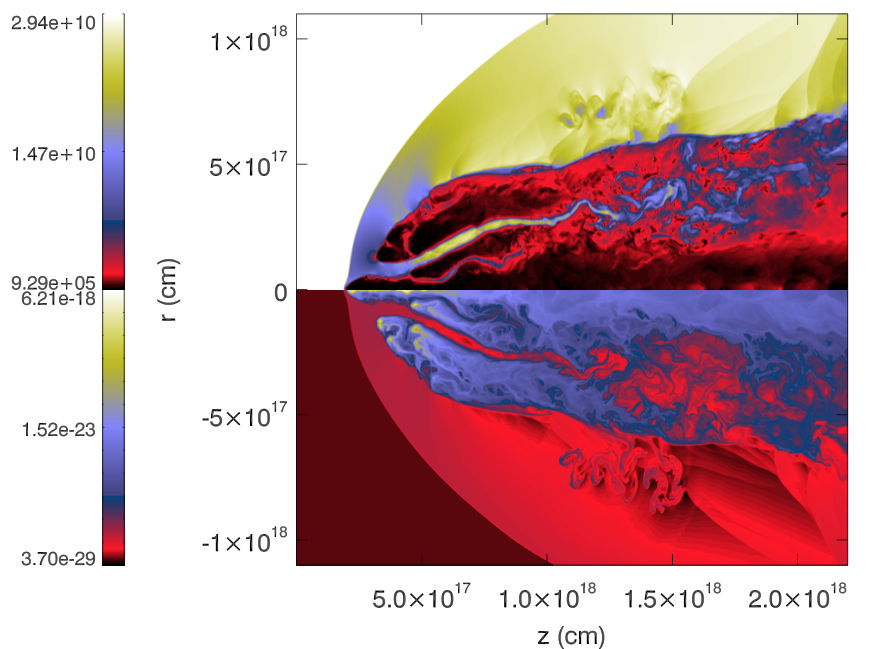}\\
  \includegraphics[clip,angle=0,width=0.96\columnwidth]{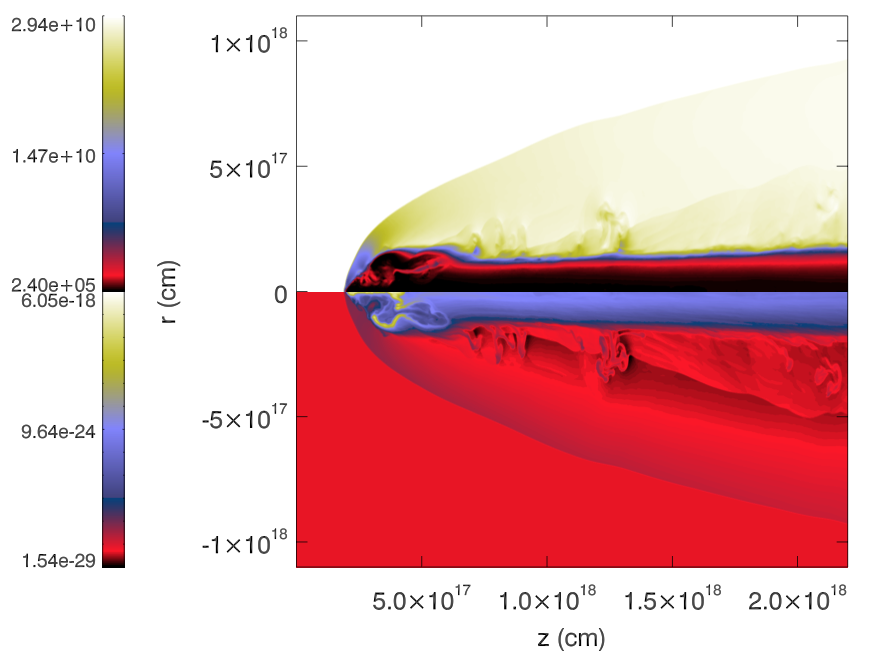}\,
\includegraphics[clip,angle=0,width=0.96\columnwidth]{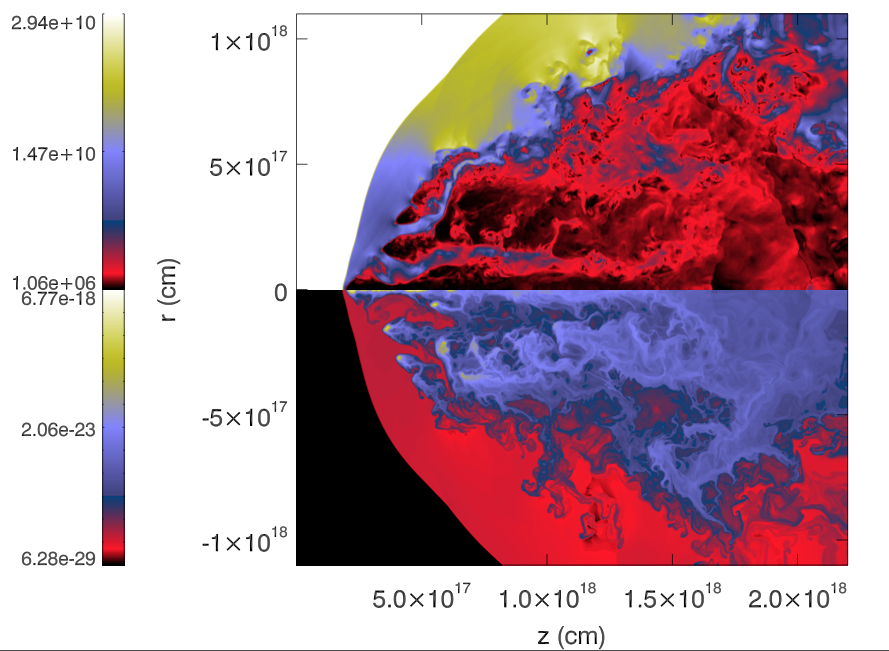}
  \caption{ Different snapshots of modulus of velocity (upper half) and rest-mass density (lower half) of simulation SW2c, in which we make profit of the axisymmetric nature of the simulation. Left column: Top panel: $t=3.6\times 10^{9}$~s. Central panel: $t=1.0\times10^{10}$~s. Bottom panel: $t=1.9\times10^{10}$~s. Right column: Top panel: $t=2.1\times10^{10}$~s. Central panel:  $t=2.2\times 10^{10}$~s. Bottom panel: $t=2.4\times10^{10}$~s.}
    \label{fig:SW2ca}
  \end{figure*} 

\begin{figure}[t]
  \includegraphics[clip,angle=0,width=\columnwidth]{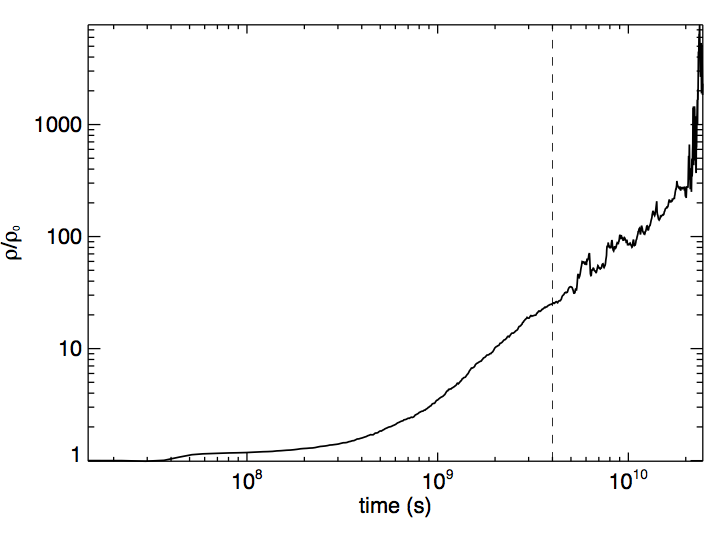}
  \caption{Normalized mean rest-mass density across the whole simulated jet cross-section for simulation SW2c, at $z=1.1\times 10^{18}\,{\rm cm}$ (coinciding with half the grid). The dashed line indicates the approximate time at which the transition to phase 2 takes place in the simulation.}
  \label{fig:rho}
  \end{figure}

Figure~\ref{fig:SW2ca} shows snapshots of simulation SW2c at different times along its evolution. We observe an initial phase (phase 1) in which the shock crosses the shell and the wind region and the mass flux slowly increases (see also Fig.~\ref{fig:rho}), and a second phase (phase 2) once the shock has crossed the shell, the bubble expands and fragments, and a large amount of material is ablated and accelerated downstream. In these series of images, we also observe that during the first phase, a cometary tail generated after the bow shock forms, with fairly homogeneous properties. This tail is close to pressure equilibrium with the shocked jet gas surrounding it and expands slowly within the shocked jet gas.

Fig.~\ref{fig:rho} shows the evolution of the mean density in the wake of the interaction within the simulated region ($z=1.1\times 10^{18}\,{\rm cm}$ from the left boundary). The value at each cell is weighted with the cell volume, which increases with radius in our axisymmetric representation. The plot shows that a single star produces a continuous increase of the mean, local jet rest-mass density during the first phase (this is expected qualitatively beyond numerical effects), and a rapid increase up to more than three orders of magnitude with respect to the initial jet density in the region at the end of the second phase. Clearly, the presence of a dense shell and the inclusion of cooling favours the local increase of density in the shocked regions and therefore, the increase of the density of the material dragged downstream. During the first phase, the jet is mass-loaded but this mass is confined to very thin, heavy, and slow jets within the jet itself. In this regard, we remark that the large amount of stars expected to fill the jet \citep[e.g.][]{ara13,wy13,mu14,bos15,wy15} would lead to a highly inhomogeneous structure with a substantial amount of its mass in the form of dense, slow, and narrow jets. Such a configuration of jets in jets does not imply a stable configuration in the long run. Regarding numerical effects, \citet{kmc94} claim that convergence in non-radiative shock-cloud interaction simulations is reached for $\sim100$ cells per cloud radius. In our simulation, we have used 64 cells per cloud radius, i.e. $\simeq 2/3$ of the suggested resolution, and thus we are not far below the convergence
resolution. Therefore, we do not expect numerical diffusion to play a crucial role. In addition, although our simulation is radiative, the cooling time is much longer than the calculation time step, meaning that cooling should not significantly affect numerical diffusion either.


  Mixing is relevant for mass-load, as it determines how the new matter is integrated in the jet flow. The axisymmetric nature of the simulation prevents the growth of the disruptive helical modes of Kelvin-Helmholtz instability \citep[see, e.g.,][and compare with the tail structure of simulation SW3 shown in next section]{pe05}. As a result, the degree of mixing of the wind gas and the jet flow is small until the end of the second phase, when the initial wind bubble is mostly carried away by the jet. Although we cannot follow mixing during the first phase in 2D simulations, we can state that the amount of gas ablated from the bubble is eventually mixed and loaded within the jet flow and use the properties of the cometary tail to study the scales in which this must happen (see section~\ref{disc}). On the contrary, the disruptive nature of the second phase favours rapid mixing within our grid. It is also relevant to mention that being the jet radius 10~pc and the star orbital velocity $1.1\times 10^7$~cm~s$^{-1}$, the crossing time of the star would be of $\sim 6\times10^{12}\,{\rm s}$. Our simulation thus shows that the initial wind bubble is completely driven away by the jet well before the star has crossed the jet.   


The final stage of the simulation resembles more the case of a homogeneous cloud than that of the inhomogeneous cloud in \cite{brpb12}. This is a consequence of the generation of an arc-shaped, cold, dense homogeneous region previous to the disruption phase (see right column in Fig.~\ref{fig:SW2ca}). The region is then destroyed like the left-overs of the homogeneous cloud in that precedent paper, after the shock had crossed the cloud. Cooling is an important qualitative difference between this stage in the simulation presented here and that of the inhomogeneous cloud in \cite{brpb12}: This process increases the density of the shocked wind region, so it plays an important role in the details of the shocked bubble evolution, enhancing also the fragmentation of the disrupted cloud.
    
\subsection{SW3}\label{res2}

  \begin{figure*}[!t]
   \includegraphics[clip,angle=0,width=0.48\textwidth]{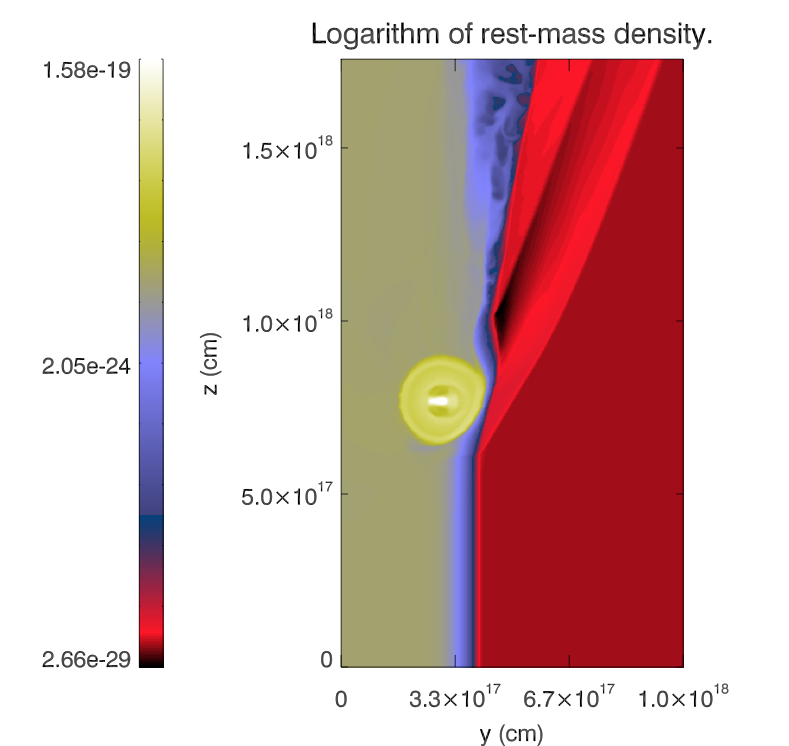}\,\,\,
  \includegraphics[clip,angle=0,width=0.48\textwidth]{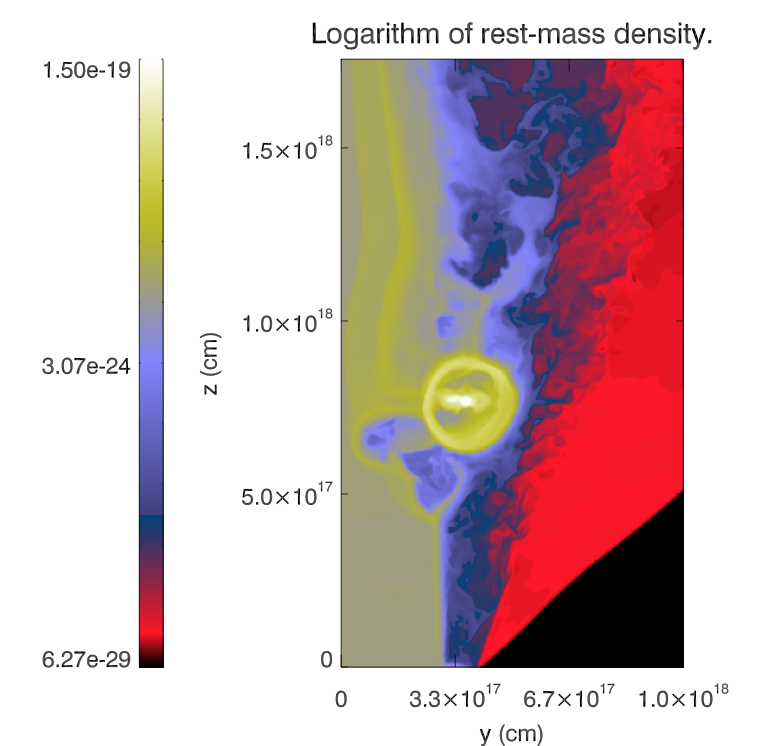}\\
  \caption{Two dimensional cuts (at $x=0\,R_0$) of rest-mass density at different times, for simulation SW3, in cgs units. The cuts are done across the center of the injector. Left panel: $t=1.3\times10^{10}\,{\rm s}$. Right panel: $t=1.8\times10^{10}\,{\rm s}$. We have cut the plotted region at $y=10^8\,{\rm cm}$ because of the lack of relevant structure farther into the jet. Note changes in the colour scales.}
  \label{fig:SW3a}
  \end{figure*}

\begin{figure*}[!t]
   \includegraphics[clip,angle=0,width=0.48\textwidth]{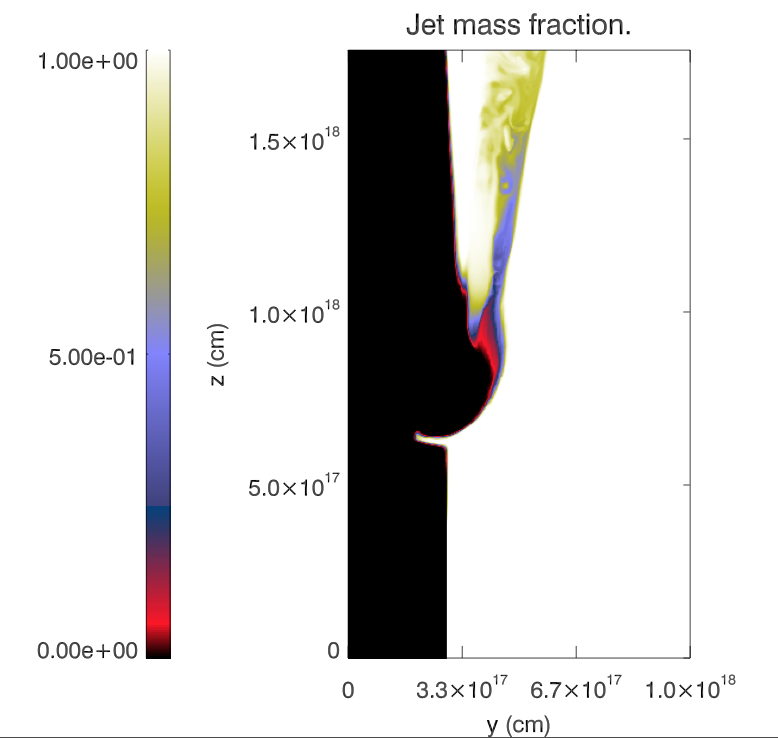} \,\,\,
  \includegraphics[clip,angle=0,width=0.48\textwidth]{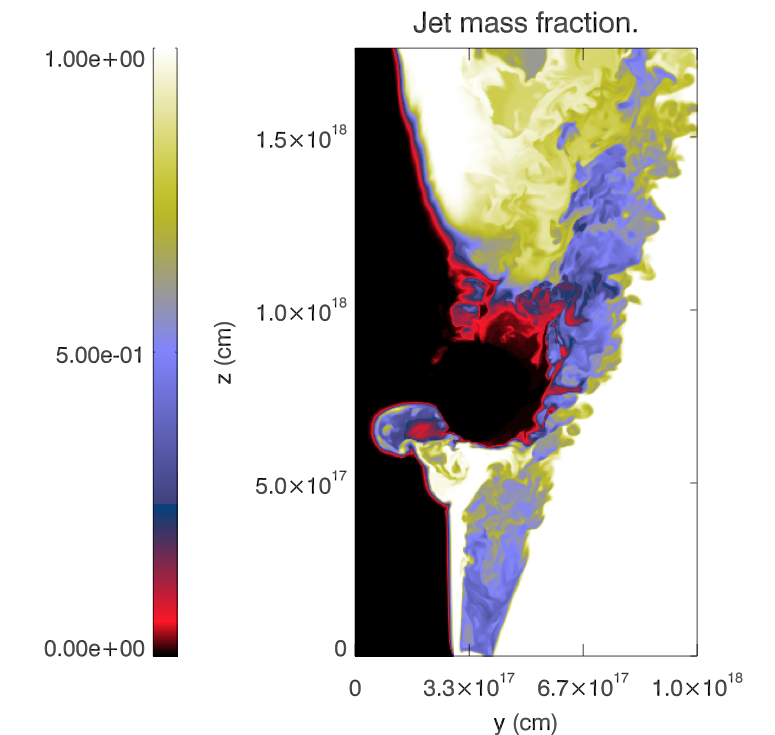}\\
  \caption{Two dimensional cuts (at $x=0\,R_0$) of jet mass fraction (tracer) at different times, for simulation SW3, in cgs units. The cuts are done across the center of the injector. The tracer is zero for the ambient medium, 1 for the jet flow, and it takes values between 0 and 1 in cells where mixing takes place. Left panel: $t=1.3\times10^{10}\,{\rm s}$. Right panel: $t=1.8\times10^{10}\,{\rm s}$. We have cut the plotted region at $y=10^8\,{\rm cm}$ because of the lack of relevant structure farther into the jet.}
  \label{fig:SW3c}
  \end{figure*}

Figures~\ref{fig:SW3a} and \ref{fig:SW3c} show cuts of the 3D grid through the centre of the injector at different times ($t=1.28\times10^{10}\,{\rm s}$, and $t=1.8\times10^{10}\,{\rm s}$) for density and tracer (the tracer is defined as zero for the ambient medium -and the bubble-, and 1 for the jet flow; it takes values between 0 and 1 in cells where mixing takes place), respectively. As the star advances towards the jet and the wind bubble touches the shear layer, it triggers a small wavelength ($<< R_j$) perturbation and a conical shock that are advected downstream, along the shear layer and into the jet, respectively (the extension of the shock is not shown beyond $y=10^{18}\,{\rm cm}$ in the right panel of Fig.~\ref{fig:SW3a}). The perturbation is shown as a double wave being advected downstream in the left panel of Fig.~\ref{fig:SW3a}. The double wave is caused by the extension of the shear layer. This short wavelength perturbation at the jet surface is not followed in our simulation, but its development can be interesting as it may trigger the growth of an instability at the jet shear layer. The wind bubble is distorted at the interaction site, showing elongation and erosion of its outer layers. The tracer plots show that strong mixing takes place at the boundary between the bubble and the jet, because of the large velocity gradient there. Another interesting feature of this initial interaction is the wave that propagates upstream along the shear layer. The axial velocity within the shear layer is small enough to allow for this upstream wave motion, unlike the case within the jet, which advects all the generated waves. The wave increases the internal energy of the region, expands the shear layer and favours upstream motion ($v_z < 0$) in the turbulent mixing region of shocked bubble and jet gas. This situation generated problems in the simulation, as the wave reached the bottom boundary of the grid and the inflow condition there was altered from that point on. This compelled us to stop the simulation. This is an issue that will be fixed in future planned simulations, and it does not affect the results presented here in the intended range of applicability.

Regarding the flow downstream of the interaction region, the conical shock enters farther into the jet flow (towards the right hand side in Figs.~\ref{fig:SW3a} and \ref{fig:SW3c}), and the shocked wind and jet material form a complex flow structure where turbulent mixing is observed already within the small grid region. The faster and hotter jet material is mixed with slower and colder shear-layer flow. A short cometary tail forms downstream of the star by wind bubble material eroded by the shear layer and jet flow. This tail is rapidly disrupted, thus favouring the mixing and acceleration of the wind gas; the mixed material visible in blue in the right panel of Fig.~\ref{fig:SW3c} propagates downstream with velocity $\simeq0.4\,c$. The tail is perturbed by the turbulent shocked jet flow and the transversal velocity of the star, which already gives transversal velocity to the tail gas.    
  
\begin{figure*}[t]
 \includegraphics[clip,angle=0,width=0.48\textwidth]{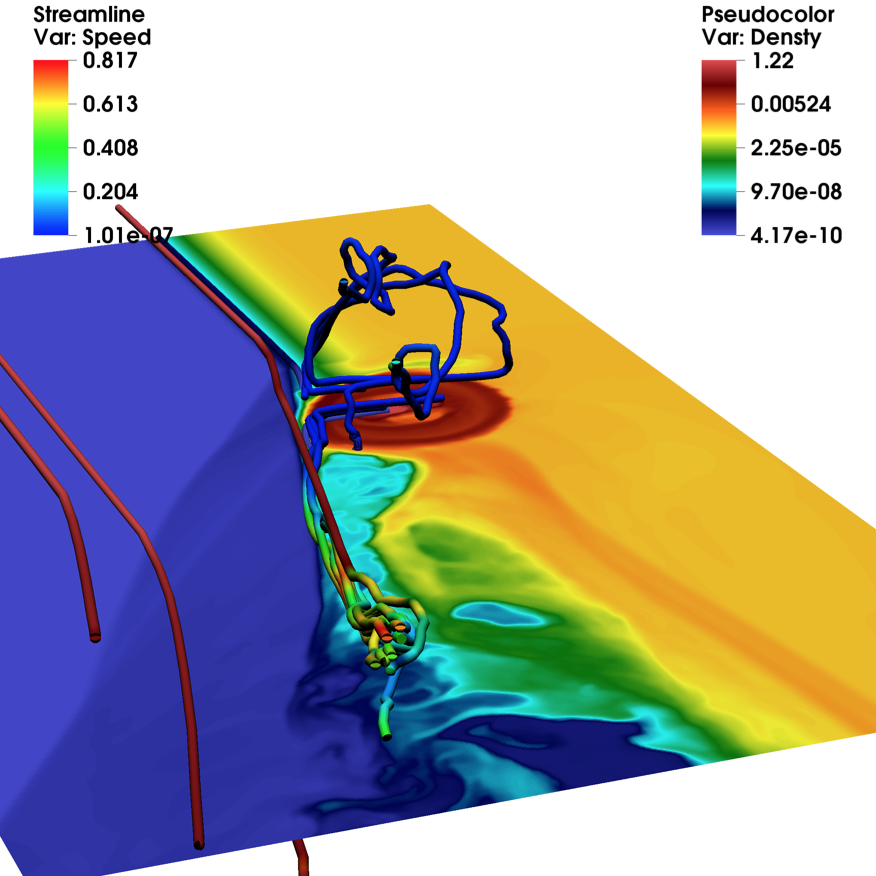} \,\,\,
  \includegraphics[clip,angle=0,width=0.48\textwidth]{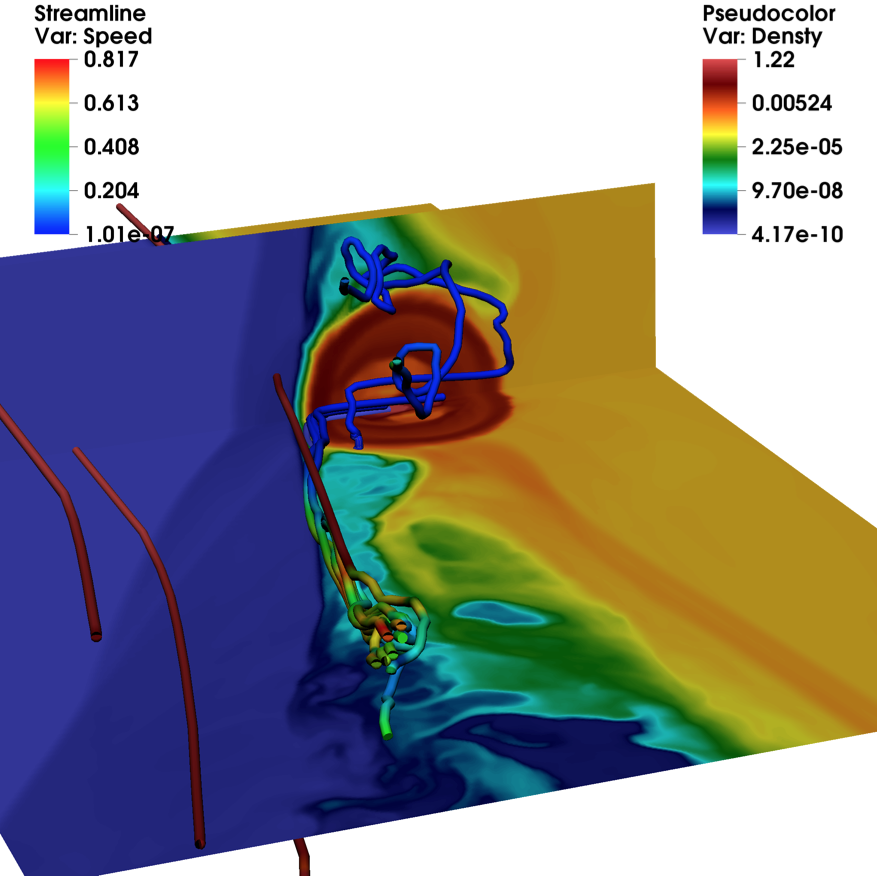}
 \caption{Three dimensional representations of rest-mass density (simulation SW3) at $t=1.4\times10^{10}\,{\rm s}$ including one cut at constant $x-$coordinate (at half grid, left panel), and the same cut plus a second one at constant $z-$coordinate (at the position of the wind injector, right panel).}
  \label{fig:SW3_3_1}
  \end{figure*}

\begin{figure*}[t]
  \begin{center}
  \includegraphics[clip,angle=0,width=0.8\textwidth]{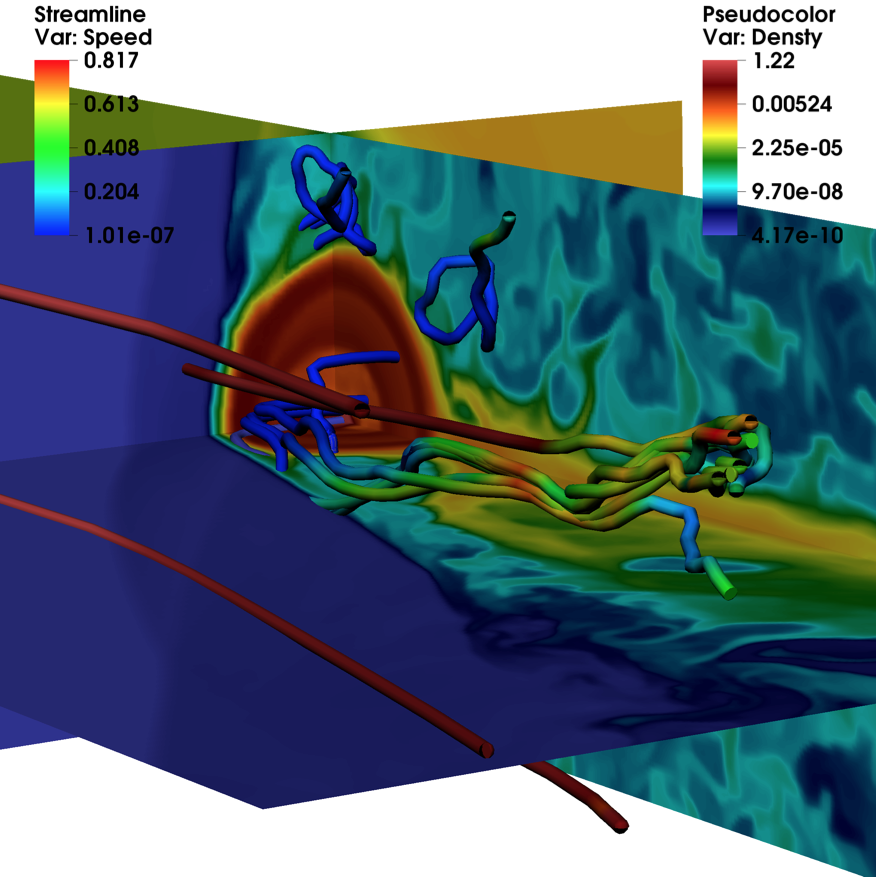}
 \caption{Three dimensional representation of rest-mass density (simulation SW3) at $t=1.4\times10^{10}\,{\rm s}$ including three cuts at half the grid size in the $x-$coordinate and at the position of the wind injector in the $z$ and $y$ coordinates.}
  \label{fig:SW3_3_2}
  \end{center}
  \end{figure*}

Figures~\ref{fig:SW3_3_1} and \ref{fig:SW3_3_2} show a projected 3D image of the rest-mass density (normalized to the wind density at the injector) at times $t=1.45\times10^{10}\,{\rm s}$. The colour scales indicate the flow speed (at the top-left of the images) and its rest-mass density (top-right). In these plots, the star enters the jet from the right of the images and propagates to the left, into the jet flow. The jet, in blue colour, propagates from the top to the bottom, on the left hand side of the images. The current lines of the jet show how the jet flow is deviated from its original direction and slightly decelerated at the site of the shock triggered by the entrance of the star. The lines starting at the wind region show the advection and tangling of the stellar wind gas downstream of the interaction region.

The tail can be seen in those plots as a well defined, light-blue region downstream of the position of the star. The images also show that the cometary tail can be easily disrupted, thus triggering efficient mixing with shocked jet material within a distance $\leq 7\times10^{16}~{\rm cm}$. Longer 3D simulations up to later stages of the interaction are planned in order to follow the star penetration into the jet and study the beginning of phase one and the eventual formation of a steady tail.
  
In summary, this simulation provides the following results: 1) The entrance of an obstacle within the jet may trigger the development of small-scale instabilities at the jet boundary and favour shear mixing when these instabilities grow to nonlinear amplitudes \citep[e.g.,][]{pe10}; 2) a conical shock is advected by the jet and propagates through its cross section, which can also trigger transitory instabilities (in the sense that the star entrance is a single event, as opposed to periodical perturbations) with longer wavelengths; 3) a shock wave propagates upstream along the shear-layer, which can lead to some energy dissipation upstream in the jet; and 4) mixing between stellar-wind tail and jet gas is very fast in 3D, owing to strong non-symmetric perturbations. The mass flux of the tail is actually affected by numerical factors such as resolution (affecting the rate at which the gas is extracted from the bubble), but we would also expect such a turbulent tail to form due to the growth of small-scale instabilities at the contact discontinuity. A numerical experiment of this (simplified) scenario in 3D is feasible, but will require future simulations with an improved setup, to avoid the shear-layer upstream wave, and a much longer running time for completion of the penetration stage.
  
\section{Discussion}\label{disc}

\subsection{Mass-loading and mixing}\label{ml2}
             
The simulations presented here show that mixing and acceleration are efficient and occur within small scales compared to the jet interaction height ($\sim 1\,{\rm pc}$ versus $100\,{\rm pc}$ in these simulations). Simulation SW2c shows that radiation cooling and the presence of a shocked wind shell surrounding the bubble lead to explosive, disruptive bubble-jet mixing in phase 2. Despite being short, simulation SW3 already shows that even within the small numerical box, i.e., at distances $\leq 10^{18}\,{\rm cm}$ mixing is hardly avoidable even in the first phase (see Figs.~\ref{fig:SW3c} and \ref{fig:SW3_3_2}). The lateral motion of the star, together with the complex interplay of waves as the star/wind bubble system enters the jet, produce nonlinear perturbations of the tail and its rapid destruction. We expect that 3D simulations including radiative losses could generate denser but still rapidly disrupting tails, able to efficiently mass-load the jet locally.

These results, together with previous studies, and in particular the unstable nature of phase 3 suggested in \citealt{del16}, indicate that the spread in the jet of the loaded matter via turbulent mixing, plus collective mass load by different stars/clouds, should globally increase the jet density and decelerate it \citep[][see also Sect.~\ref{sec:swi}]{wy13,pmlh14,wy15}. The actual mass loaded in the jet, i.e. the actual stellar mass-loss rates and the collective effects of many stars, will determine whether this process is important for the jet content \citep[e.g.,][]{pmlh14}. Although efficient mixing is predicted, the mass-loading process may still lead to strong jet inhomogeneities in density and composition, if the loaded material is not rapidly spread across the jet cross-section. 

We have computed the amount of mass dragged by the jet after $\sim 10^{10}\,{\rm s} \simeq 10^3\,{\rm yr}$, the dynamical time of the bubble, and it is $\approx 2\times\,10^{-2}\,M_\odot$, which is compatible with the adopted total mass of the bubble plus the dynamical time and plus mass-loss rate of the star. Making the raw assumption that the loaded mass is homogeneous in time (although it is actually concentrated towards the end of phase 2, see Fig.~\ref{fig:rho}), we obtain a mean input $\sim 10^{-5}\,M_\odot/{\rm yr}$ for a single AGB star during phases one and two (and in phase three as well, as wind keeps being injected in the jet). Such a mass-load rate is of the order of the mass flux expected in FRI jets \citep[see, e.g.,][]{lb02b,PM07}. This is therefore a critical stage in the process of mass-loading by stars and should be taken into account in future numerical simulations of jet deceleration \citep{pmlh14}.
Interestingly, $\sim 0.01$\% of $\sim 1\,M_\odot$-stars are at present AGB stars. Thus, the jet of a radiogalaxy such as M87 may contain $\sim 10^3$ AGB stars up to kpc scales \citep[e.g. see the discussion in][]{bos15}. Note nevertheless that this conclusion applies to mass-load when the bubble is well inside the jet, but the mass-load may be also relevant during the early jet-penetration stage of the bubble, when it is in direct interaction with the jet boundary.

\subsubsection{Mass-load in the shear layer}

In this work, we have assumed that, once the bubble is inside the jet, the bubble radius ($R_{b}$) is determined by equating the time needed by the bubble to enter the jet to the time needed for the jet to shock the bubble. However, the radius of the bubble before entering the jet is determined by equating the wind and the medium (ram) pressures. This yields a value for $R_b$ outside the jet larger by a factor $\sim \max \left(\, (P_{jet}/P_{ISM})^{1/2}\,(v_w/v_{orb}), \,1\right)$ than inside the jet, with $P_{ISM}=\rho_{ISM} v_{orb}^2$ (where $v_{orb}$ is the star velocity). The consequence of this is that the mass-load in the jet shear layer will be also larger by the same factor. Given that SW3 shows that the matter loaded in the shear layer efficiently spreads inside the jet, it seems reasonable to assume that all the mass accumulated in the bubble before jet penetration is eventually entrained by the jet but when $R_b\gtrsim R_j(z)$, i.e., close enough to the black hole. Upstream of this point, the specific bubble-jet contact geometry should be taken into account.

In fact, the bubble radius outside the jet may be so large that the mass injected by the bubble to the jet shear layer surpassed that injected in phase 3. This would happen for $R_{b}\gtrsim R_{j}\,(v_{\rm w}/v_{\rm orb})$, which will occur at:
\begin{equation}
z\lesssim 10\,\frac{\dot{M}^{1/2}_{w,-5}}{n^{1/2}_0v^{1/2}_{w,6.5}\xi_{-1}}\,{\rm pc}\,,
\end{equation}
where $n_0$ is the number density of the ambient medium (assuming hydrogen), $\xi$ is the jet opening angle (taken to be 0.1 radians in the paper, see section~\ref{sim}), and we have adopted the quantity format $A_x=A/10^x$, with $A$ in the corresponding units.

We thus conclude from these basic estimates that first, shear-layer entrainment that may dominate mass-load occurs during the early jet penetration stage. Then, this work shows that phases 1-2 lead to an explosive release of mass, smaller than in the penetration phase and well inside the jet. And finally, phase 3 mass-load occurs more smoothly than phase 2 but still in an unstable manner all through the jet \citep{del16}.

\subsubsection{Scalability of the results}

The star mass-loss rate adopted here is high, more typical for an AGB than a red giant. Nevertheless, we note that for the 3D, adiabatic simulation SW3, the results can be scaled keeping $\dot{M}/L_j$ constant, where $L_j$ is the jet kinetic power. In addition, the general discussion on stability should qualitatively apply as well to lower mass-loss rate stars for the same $L_j$, although a quantitative assessment requires specific simulations, and the same applies to faster winds, such as those of massive stars. Simulation SW2c is not scalable, as it includes more detailed physics that require fixing units. Lighter winds could still cool down effectively, given the dynamical time-scales of the simulations, as the shell cooling time-scales are short enough; this is safely the case for a range of $\dot{M}$ spanning few orders of magnitude below the adopted value. This can be seen estimating the cooling effect on the region of the stellar wind shock. The post-shock temperature should be $\approx 4\times10^4$~K, which implies a very fast cooling regime. The cooling rate is $Q=\lambda n^2$, where $\lambda = 10^{-22}\,\mbox{erg cm}^3~$s$^{-1}$. The wind density, without compression, is $10^5\, {\rm cm}^{-3}$, so the cooling time would be already $t_{\rm cool}\approx 10^6$~s; accounting for shock compression, $t_{\rm cool}\approx 2\times 10^5$~s. These time-scales should be compared with the adiabatic evolution time, $t_{\rm add}\sim R_s/v_w\approx 3\times 10^9$~s, and to the simulation time-step ($\simeq 1.6\times10^4$~s). Thus, for SW2c, one has $t_{\rm add}\gg t_{\rm cool}$. Although wind cooling is thus of particular importance for RG/AGB stars, it must be taken into account even for the case of lighter stellar winds. Regarding massive stars, their winds are less dense and much faster, and therefore the cooling will be significant only for much closer shock radii, i.e. for very powerful jets, or much closer to the jet base.

\subsubsection{Tail stability and mixing scales}

We focus now on the stability properties of wind-tails, a relevant aspect to mixing. In 3D, the motion of the star across the jet already provides the tail with a transversal velocity, inducing the growth of helical structures. This perturbation is nonlinear and favours mixing locally or, at least, close to the interaction region comparing to the jet scale. On the contrary, the tail of shocked wind material in 2D appears to generate a stable structure as the shock crosses the bubble (first phase). The axisymmetry and the absence of any perturbation in a region that is in pressure equilibrium with its surroundings avoids the trigger of disruptive instability modes. Despite these simplifications, the properties of the tail obtained in the simulation are of the same order as those for the 3D case: the tail shows densities of $10^{3}-10^{4}$ times the shocked jet density and typical velocities $\simeq 0.05-0.1\,c$ during this transitory phase that lasts for $\sim 10^{10}\,{\rm s}$. Assuming that the mass flux is indeed mimicking an actual physical process despite numerical contamination (see Sect.~\ref{res2}), we took these typical values to run a stability analysis of the tail/shocked jet system. 
   
We calculated the solutions to the stability problem of a sheared flow \citep[see, e.g.,][]{pe07}. The formation of a shear layer is expected in situations in which there is some mixing-dissipation at the contact discontinuity. The formation of the tail from the erosion of the external layers of the bubble should lead to such structure. We adopted the spatial approach in our stability calculation, so we obtained the real and imaginary parts of the wavenumber component along the $z$-direction, $k_{z,r}$ and $k_{z,i}$, respectively, in terms of a real value of the frequency $\omega$ for solutions of the type $\delta(r,\theta,z,t)= A_0  \,e^{-i(k_r r + n \theta + (k_{z,r}+ i k_{z,i}) z+\omega t)} = A_0 \,e^{k_{z,i} z} \,e^{-i(k_r r + n \theta + k_{z,r} z+\omega t)}$, where $k_r$ is the radial wavenumber (a function of $k_z$ and $\omega$), and $n=0,1$ for the pinching and helical modes, respectively \citep[see, e.g.,][]{pe05}. Neglecting the imaginary part of the radial wavenumber, $k_r$, with respect to the imaginary part of the axial wavenumber, $k_z$, we can write the axial dependence of the amplitude as $A(z)= A_0\,e^{k_{z,i} z}$. The units are given by the speed of light and the radius of the tail ($R_t\simeq 10^{17}\,{\rm cm}$).

 \begin{figure*}[!t]
  \includegraphics[clip,angle=0,width=0.48\textwidth]{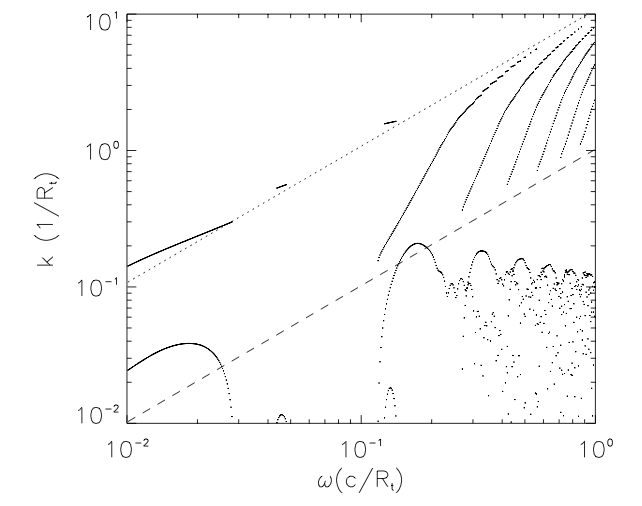}
  \includegraphics[clip,angle=0,width=0.48\textwidth]{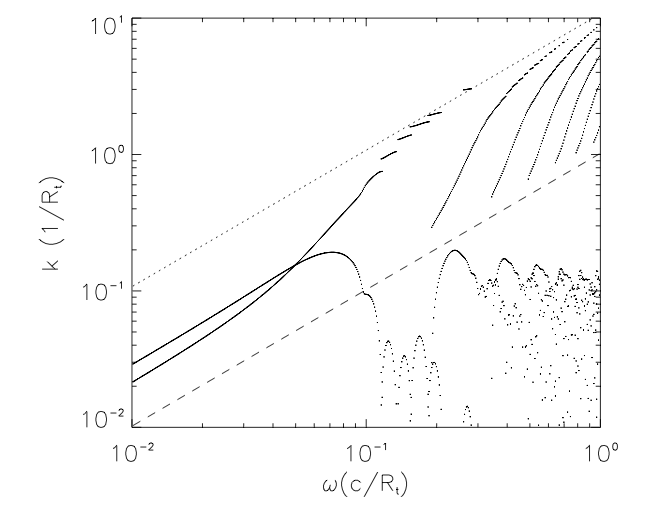}
  \caption{Solutions of the linear perturbation equation for typical profiles of rest-mass density and axial velocity obtained from the simulations (left panel: pinching, symmetric mode; right panel: helical mode). The tail density is taken as $\rho_t/\rho_j = 10^3$, the tail and jet velocities are $v_t= 0.05\,c$ and $v_j=0.95\,c$, respectively, and the sound speeds are $c_{s,t}=0.04\,c$ and $c_{s,j}=0.54\,c$. The dashed and dotted lines show the line corresponding to phase velocities, $w/k$, equal to the relativistic composition of the velocity and sound speed for each of the flows, i.e., $0.98\,c$ for the jet and $0.09\,c$ for the tail. These velocities separate the solution plane in different allowed and forbidden regions \citep[see, e.g.,][for the non-relativistic case]{pc85}.}
 \label{fig:KH1}
  \end{figure*}

The solutions to the stability problem are displayed in Fig.~\ref{fig:KH1}. The left panel shows the solutions for the pinching mode ($n=0$), and the right panel shows the solutions for the helical mode ($n=1$), for the stability equation using $\rho_t/\rho_j = 10^3$ (in the figure; $\rho_t/\rho_j = 10^4$ has also been calculated, but is not shown), $v_t= 0.05\,c$ and $v_j=0.95\,c$, and specific internal energies of $\epsilon_t=2.5\times10^{-3}\,c^2$ and $\epsilon_j= c^2$, which are typical values given by our simulations. The solutions have been derived for the case of a transition with a thickness of the same order as the tail radius $R_t$, as given also by the simulations. Each of the curves observed in the panels displayed in Fig.~\ref{fig:KH1} stand for a particular unstable mode that can be triggered with different frequencies and wavelengths: the left panel shows the solutions for the pinching, symmetric mode, whereas the right panel shows the solutions for the helical mode. The upper curves within each panel stand for $k_{z,r}$, whereas the lower ones represent the corresponding $k_{z,i}$. The inverse of the latter is known as the growth length, which defines the distance at which the amplitude of the perturbation e-folds. The solutions were obtained point to point using the shooting method \citep[a combination of a Runge-Kutta integrator with variable step and a root finding M\"uller method,][]{rcl84,pe07}, which is the reason for the discontinuous appearance of the curves (this is irrelevant for our analysis, though). The denser tail ($\rho_t/\rho_j = 10^4$) results in slightly longer, but similar, growth lengths (smaller values of $k_{z,i}$). Increasing the size of the layer would increase the growth distances, i.e., stabilize, of short wavelength modes ($\lambda \leq R_{sh}$, with $R_{sh}$ the size of the transition), whereas the properties of long wavelength modes would not change significantly \citep[e.g.,][]{pe05}.

In the case of the pinching mode, the solutions of the fundamental mode (a mode with no nodes between the tail axis and the shear layer of the tail) appear only at small phase speeds ($k>>w$) and show small relative growth rates. The helical surface mode is, on the contrary, the most unstable for low frequencies (long wavelengths). Body modes are characterized by their different radial structure and they appear as different curves in the solution, higher order modes showing up as $k_r$ grows.     
The minimum growth lengths obtained from the solutions of the linear problem correspond to maximum values of $k_i^* \simeq 2\times 10^{-1} R_t^{-1}$. Considering a perturbation of a ten per cent of the background value, $A_0=0.1$ (the amplitude is normalized to the background value) for variable $X$, the perturbation will become fully nonlinear when $A\sim 1$. Taking into account that $A(z)=A_0\,e^{k_i z}$, and using the $k_i^* \simeq 10^{-1} R_t^{-1}$, $A$ becomes $\sim1$ at $z\simeq 20\,R_t \simeq 10^{18}-10^{19}\,{\rm cm}$, for $R_t = 5\times10^{16} - 10^{17}\,{\rm cm}$. This means that the tail material is expected to mix with the jet within distances $\sim 1\,{\rm pc}$ if an initially stable tail was perturbed. From previous stability studies \citep[e.g.,][]{pe05} we can claim that these distances are reduced when the relative density of the tail with respect to the jet decreases (i.e., closer to the active nucleus), and that they are longer when this relative density increases (e.g., towards the end of phase two). The increase of the mass-flux at the end of the second phase could result in an important increase of the growth lengths and the result in the formation of longer tails. Nevertheless, if our estimates are correct, one would expect to find inhomogeneous jet structure caused by tails $\sim 1\,{\rm pc}$ long downstream of jet/star interaction sites. Such a scenario was recently suggested for VLBI observations of the jet in Centaurus~A by \citet{mu14}. We can conclude that tails are, in principle, stable transient structures due to their inertia \citep{pe05}, but that the strong perturbations expected, and observed in simulation SW3, may force efficient mixing relatively close to the interaction site ($\simeq 1\,{\rm pc}$). 

It is also relevant to mention that the explosive nature of the disruption of the wind bubble will contribute to efficient mixing on the scales of the interaction region. Such explosive events may be bright enough to be observed in close sources. If we consider a mean jet radius of $R_j\sim 10\,{\rm pc}$ along its trajectory within the host galaxy and a mean orbital velocity of $v_o \sim 10^7$~cm~s$^{-1}$, the mean crossing time of the stars is $t_c\sim 6\times 10^{12}\,{\rm s}$. Taking into account that the disruption phase lasts for $\sim 10^{10}\,{\rm s}$ (see Fig.~\ref{fig:rho}), we expect that about one thousandth of the red-giant stars in the jet should be at this stage and possibly produce bright spots within the jet as those observed in Centaurus A \citep{wo08,go10}.

\subsection{Non-thermal activity}

If particles are accelerated at the jet shock, recent numerical studies of cloud/stellar wind-jet interactions, and of their related non-thermal emission, show that the interaction region will produce significant amounts of detectable radiation \citep{bos15,del16,vtb17}. The main factors behind the high apparent radiation efficiency are 1) jet energy dissipation, which actually occurs on scales much larger than the jet shock stand-off distance from the obstacle \citep{bos15}, and 2) Doppler boosting effects caused by a post-shock flow that is also moving at relativistic speeds \citep[see also the analytical studies of][in which the obstacle itself moves at relativistic velocities]{ba12b,kh13}. Confirming and extending the results of \cite{brpb12} for a homogeneous obstacle, the present work suggests that the initial wind bubble can also trigger substantial non-thermal activity. Given its size, much larger than the more-or-less steady interaction structure simulated for instance by \citet{bos15} and \citet{del16}, the bubble represents a temporal target for the jet that may eventually cover a significant fraction of the jet section \citep[as in][]{kh13}. The details of the subsequent radiation will strongly depend on the height of the interaction region, in particular the balance between radiative and non-radiative losses \citep{bos15}, and in certain cases the bubble penetration into the jet may lead to detectable transient emission. A quantitative study of this emission is out of the scope of this work, but it is worth noting that the star penetration stage should be also accounted for when assessing the role of stars in jet high-energy emission. We also remark that such a study would also strongly benefit from the inclusion of the magnetic field in the simulations.

\subsection{Stellar wind induction}
\label{sec:swi}
 
If non-thermal particles are produced, their presence can increase the wind mass loss rate if the jet shock can reach close to the stellar surface. This will happen if the jet has a very high ram pressure, or when the interaction takes place close to the jet base \citep{kh13}. This mass loss rate amplification is close to the well-known effect of wind induction by X-ray radiation in close X-ray binaries and in AGNs \citep{bs73,dor08a,dor08b}. We estimate in what follows the wind mass-loss rate following \cite{kh13}. 

One can estimate the heating rate due to non-thermal particles interacting with the stellar photosphere as $F_{nt}= \chi L_j/\pi R_j^2$, where $\chi$ is the efficiency of particle acceleration. The excited mass flux can be estimated as $\mu = 10^{-12} \alpha_{-12} F_{nt} R_{*}^{1/2}/(G M_{*})^{1/2}$~g/s~cm$^2$, where $\alpha_{-12}$ is the efficiency of wind induction, and
$R_*$ and $M_*$ the stellar radius and mass, respectively. The parameter $\alpha_{-12}$ is $\simeq 1$ in the case of X-ray heating; we take it here as a fiducial value but it should be derived for the specific case of non-thermal heating. The total mass loss rate in the wind can be estimated as
\begin{equation}
 \dot{M}=\mu \pi R_*^2 = 10^{-12}\alpha_{-12}\chi L_j \frac{R_*^2}{R_j^2} \frac{ R_*^{1/2}}{\left(G M_*\right)^{1/2}}.
 \label{eq:midot}
\end{equation}

To compare the intensity of the induced wind and the normal mass-loss rate, we remind that the mass-loss rate of RG/AGB star is assumed to be $\dot{M}=10^{-5}M_{\odot}$/year 
or $0.7\times10^{21}$~g/s. Combining this value with Eq.~\ref{eq:midot}, we get the condition under which the induced stellar wind starts to dominate over the normal one:
\begin{equation}
    R_j < 100 R_* \chi^{1/2}\alpha_{-12}^{1/2} L_{j,44}^{1/2} \upsilon_{c,7}^{-1/2},
 \label{eq:mdc}
\end{equation}
where $\upsilon_c = \sqrt{G M_*/R_*}$. 

A jet can be significantly decelerated by the wind if its power is smaller than the critical value $L_{\rm crit} = \dot{M} c^2 \Gamma_j = 6\times 10^{43}  \dot{M}_{-4}  \Gamma_{j,1}  $~erg~s$^{-1}$.
In the induced wind case, the jet deceleration limit does not depend on jet power, and one can estimate the maximum jet radius for which the jet will be decelerated by the wind as
\begin{equation}
  R_j< 10 \Gamma^{1/2} N_*^{1/2}  \chi^{1/2}\alpha_{-12}^{1/2} \upsilon_{c,7}^{-1/2} R_*,
 \label{eq:om}
\end{equation}
where $N_*$ is number of stars located in the jet at a given distance from the SMBH. In conclusion, the induced stellar wind could be a dynamically important factor for stars with big radius passing the jet close to its base (not simulated here) and independently of its power.
      
\subsection{The potential effects of magnetic field}

A further step of this study should be performed including magnetic fields in order to derive more conclusive results, taking into account that the magnetic field can be a relevant actor at the interaction region. Moreover, a gas cloud embedded in a jet, like in the scenario studied in this paper, can trigger the growth of Rayleigh-Taylor instability \citep{cha61}. In this respect, the magnetic field can also play an important role \citep{ims72}. A brief summary of the role of magnetic field is in order here. 

Following seminal papers by \cite{kmc94} and \cite{pfb02} on the interaction between clouds and shocks in a hydrodynamical context, several works have studied the influence of magnetic fields on the interaction between gas clouds and shocks in the context of supernova remnants or dense clouds interacting with the interstellar medium \citep[see, e.g.,][]{ml94,jrt96,min99,gre99}. In \citet{ml94} and \citet{jrt96} the authors study the effect of parallel and perpendicular fields on clumps of gas interacting with winds using axisymmetric simulations mainly. These works indicate that the orientation of the field has relevant consequences for the interaction. On one hand, when the magnetic field is parallel to the shock surface, it is amplified by the interaction and, when stretched around the cloud, 
it prevents its disruption in that direction, but can amplify cloud disruption in the axial direction, thus leading to asymmetrical cloud disruption. On the other hand, when the field is perpendicular to the shock (axial), the lines are stretched around the cloud and show reversals that may induce reconnection in this region. The magnetic tension of the axial field prevents the formation of vortexes, also reducing the fragmentation of the cloud. For oblique fields, the evolution resembles more to one case or the other depending on the angle between the lines and the direction of motion \citep{jrt96,min99}. \citet{gre99} presented 3D simulations in which the magnetic field is perpendicular to the propagation, deriving similar results. To our knowledge, there are still no works that study the evolution of the tail, although in \citet{min99} the asymmetric nature of the simulations allowed oscillations in the tail that could couple to a kink mode of the current-driven instability or a helical KH mode.

\section{Summary}\label{sum}

We have performed 2D and 3D simulations of the first stages of the interaction between a star surrounded by its wind bubble and an AGN jet. At the beginning of the interaction, the wind bubble is shocked while being slowly eroded by the shocked jet flow. This leads to a transient continuous cometary-like tail of wind material, which likely gets disrupted locally, as 3D simulations and analytical stability calculations show. This phase is followed by the eventual expansion, disruption, and spread of the material initially present in the wind bubble. For this to occur, the presence of a previous shocked wind shell and radiation cooling are important, showing that the case of a $r^{-2}$-density profile wind actually resembles that of a homogeneous cloud.
Although steady jet-wind interaction when the two flows balance each other pressures has not been simulated here, complementary results from \cite{del16} show that the interaction structure is also likely unstable, which would favour jet-wind mixing. Finally, despite the jet-wind interaction dynamics tends to distribute the stellar wind mass on large jet regions, some degree of inhomogeneity may be expected in extragalactic jets. The mass-loss rate adopted, $10^{-5}\,\,M_\odot/{\rm yr}$, which is also the mean mass-load in the jet, represents a significant amount of the expected mass fluxes in FRI jets. In fact, the mass entrained through the shear layer could be a substantial fraction of, or even dominate, the total mass loaded by a star into the jet.
If a moderate number of AGB stars are present in the centre of galaxies, as expected, then AGN stars could easily be important channels for mass-loading in AGN jets.

The transitory wind bubble penetration of the jet should be studied as a potential source of non-thermal emission, in addition to other stages of the jet-star interaction, and the associated energetic particles may induce additional jet mass-load for jet-star interactions close to the stellar atmosphere. The magnetic field must be included in order to more accurately characterize jet-wind mixing, and the scenario-related non-thermal processes. 

\begin{acknowledgements}
The authors acknowledge Sarka Wykes for comments on the manuscript and discussion. MP acknowledges support by the Spanish ``Ministerio de Econom\'{\i}a y Competitividad'' grants AYA2013-40979-P, and AYA2013-48226-C3-2-P. The research leading to these results has received funding from the European Union Seventh Framework Program (FP7/2007-2013) under grant agreement PIEF-GA-2009-252463. V.B-R. acknowledges support by the Spanish Ministerio de Econom\'{i}a y Competitividad (MINECO/FEDER, UE) under grants AYA2013-47447-C3-1-P and AYA2016-76012-C3-1-P with partial support by the European Regional Development Fund (ERDF/FEDER), MDM-2014-0369 of ICCUB (Unidad de Excelencia `Mar\'{i}a de Maeztu'), and the Catalan DEC grant 2014 SGR 86. V.B.R. also acknowledges financial support from MINECO and European Social Funds through a Ram\'on y Cajal fellowship. This research has been supported by the Marie Curie Career Integration Grant 321520. BMV acknowledge the partial support by the NSF grant AST-1306672 and DoE grant DE-SC0016369. The authors thankfully acknowledge the computer resources, technical expertise and assistance provided by the ''Centre de C\`alcul de la Universitat de Val\`encia'' through the use of Tirant, the local node of the Spanish Supercomputation Network, and that also provided by the Barcelona Supercomputing Center - ''Centro Nacional de Supercomputaci\'on''.
\end{acknowledgements}

\end{document}